\documentclass[a4paper]{article}
\pdfoutput=1

\usepackage{lpic}
\usepackage{INTERSPEECH2018}
\usepackage{multirow}
\usepackage{graphicx}
\usepackage{subcaption}
\usepackage{microtype}
\usepackage{amsmath,amssymb}
\usepackage{nicefrac}

\title{Single headed attention based sequence-to-sequence model \\ for state-of-the-art results on Switchboard}
\name{Zolt{\'a}n T{\"u}ske, George Saon, Kartik Audhkhasi\sthanks{\ \ \scriptsize currently at Google Inc.}, Brian Kingsbury}
\address{IBM Research AI, Yorktown Heights, USA}
\email{zoltan.tuske@ibm.com}

\begin{document}

\renewcommand{\baselinestretch}{0.933}\normalsize
\maketitle

\begin{abstract}
It is generally believed that direct sequence-to-sequence (seq2seq) speech recognition models are competitive with hybrid models only when a large amount of data, at least a thousand hours, is available for training.
In this paper, we show that state-of-the-art recognition performance can be achieved on the Switchboard-300 database using a single headed attention, LSTM based model.
Using a cross-utterance language model, our single-pass speaker independent system reaches 6.4\% and 12.5\% word error rate (WER) on the Switchboard and CallHome subsets of Hub5'00, without a pronunciation lexicon.
While careful regularization and data augmentation are crucial in achieving this level of performance, experiments on Switchboard-2000 show that nothing is more useful than more data.
Overall, the combination of various regularizations and a simple but fairly large model results in a new state of the art, 4.7\% and 7.8\% WER on the Switchboard and CallHome sets, using SWB-2000 without any external data resources.

\end{abstract}

\section{Introduction}
\label{sec:intro}
Powerful neural networks have enabled the use of ``end-to-end'' speech recognition models that directly map a sequence of acoustic features to a sequence of words without conditional independence assumptions.
Typical examples are attention based encoder-decoder~\cite{Bahdanau2015} and recurrent neural network transducer models~\cite{Graves2012}.
Due to training on full sequences, an utterance corresponds to a single observation from the view point of these models; thus, data sparsity is a general challenge for such approaches, and it is believed that these models are effective only when sufficient training data is available.
Indeed, many end-to-end speech recognition papers focus on LibriSpeech, which has 960 hours of training audio.
Nevertheless, the best performing systems follow the traditional hybrid approach~\cite{Bourlard1993}, outperforming attention based encoder-decoder models~\cite{Park2019,Karita2019,Luscher2019,Wang2019}, and when less training data is used, the gap between ``end-to-end'' and hybrid models is more prominent~\cite{Park2019,Irie2019asru}.
Several methods have been proposed to tackle data sparsity and overfitting problems; a detailed list can be found in Sec.~\ref{sec:methods}.
Recently, increasingly complex attention mechanisms have been proposed to improve seq2seq model performance, including stacking self and regular attention layers and using multiple attention heads in the encoder and decoder~\cite{Karita2019,Vaswani2017}.

We show that consistent application of various regularization techniques brings a simple, single-head LSTM attention based encoder-decoder model to state-of-the-art performance on Switchboard-300 (SWB-300), a task where data sparsity is more severe than LibriSpeech.
We also note that remarkable performance has been achieved with single-head LSTM models in a recent study on language modeling \cite{Merity2019}.

\renewcommand{\baselinestretch}{0.941}\normalsize
\section{Methods to improve seq2seq models}
\label{sec:methods}
In contrast to traditional hybrid models, where even recurrent networks are trained on randomized, aligned chunks of labels and features~\cite{Saon2014,Mohamed2015}, whole sequence models are more prone to memorizing the training samples.
In order to improve generalization, many of the methods we investigate introduce additional noise, either directly or indirectly, to stochastic gradient descent (SGD) training to avoid narrow, local optima.
The other techniques we study address the highly non-convex nature of training neural networks, ease the optimization process, and speed up convergence.
\\ {\bf Weight decay} adds the $l_2$ norm of the trainable parameters to the loss function, which encourages the weights to stay small unless necessary, and is one of the oldest techniques to improve neural network generalization.
As shown in \cite{Krogh1992}, weight decay can improve generalization by suppressing some of the effects of static noise on the targets.

\noindent {\bf Dropout} randomly deactivates neurons with a predefined probability in every training step~\cite{hinton2012} to reduce co-adaptation of neurons.

\noindent {\bf DropConnect}, which is similar in spirit to dropout, randomly deactivates connections between neurons by temporarily zeroing out weights~\cite{pmlr-v28-wan13}.

\noindent {\bf Zoneout}, which is also inspired by dropout and was especially developed for recurrent models~\cite{Krueger2017}, stochastically forces some hidden units to maintain their previous values.
In LSTMs, the method is applied on the cell state or on the recurrent feedback of the output.

\noindent {\bf Label smoothing} interpolates the hard label targets with a uniform distribution over targets, and improves generalization in many classification tasks~\cite{Szegedy2016}.

\noindent {\bf Batch normalization (BN)} accelerates training by standardizing the distribution of each layer's input~\cite{Ioffe2015}.
In order to reduce the normalization mismatch between training and testing, we modify the original approach by freezing the batch normalization layers in the middle of the training when the magnitude of parameter updates is small.
After freezing, the running statistics are not updated, batch statistics are ignored, and BN layers approximately operate as global normalization.

\noindent {\bf Scheduled sampling} stochastically uses the token produced by a sequence model instead of the true previous token during training to mitigate the effects of exposure bias~\cite{Bengio2015}.

\noindent {\bf Residual networks} address the problem of vanishing and exploding gradients by including skip connections~\cite{He2016} in the model that force the neural network to learn a residual mapping function using a stack of layers.
Optimization of this residual mapping is easier, allowing the use of much deeper structures.

\noindent {\bf Curriculum learning} simplifies deep neural network training by presenting training examples in a meaningful order, usually by increasing order of difficulty~\cite{Bengio2009}.
In seq2seq models, the input acoustic sequences are frequently sorted in order of increasing length~\cite{Amodei2016}.

\noindent {\bf Speed and tempo perturbation} changes the rate of speech, typically by $\pm$10\%, with or without altering the pitch and timbre of the speech signal~\cite{Kanda2013,Ko15}.
The goal of these methods is to increase the amount of training data for the model.
\\{\bf Sequence noise injection} adds structured sequence level noise generated from speech utterances to training examples to improve the generalization of seq2seq models~\cite{Saon2019}.
As previously shown, input noise during neural network training encourages convergence to a local optimum with lower curvature, which indicates better generalization~\cite{bishop95}.
\\{\bf Weight noise} adds noise directly to the network parameters to improve generalization~\cite{Murray1994}.
This form of noise can be interpreted as a simplified form of Bayesian inference that optimizes a minimum description length loss~\cite{Graves2011}.
\\{\bf SpecAugment} masks blocks of frequency channels and blocks of time steps~\cite{Park2019} and also warps the spectrogram along the time axis to perform data augmentation.
 It is closely related to~\cite{Zhong2017}.

\section{Experimental setup}
This study focuses on Switchboard-300, a standard 300-hour English conversational speech recognition task.
Our acoustic and text data preparation follows the Kaldi~\cite{Povey_ASRU2011} \texttt{s5c} recipe, which is based on the transcription release of Mississippi State University \cite{msu}.
Our attention based seq2seq model is similar to \cite{Bahdanau2016,Chan2016} and follows the structure of \cite{Tuske2019}.

We extract 80-dimensional log-Mel filterbank features over 25ms frames every 10ms from the input speech signal.
The input audio is speed and/or tempo perturbed with {\large \nicefrac{5}{6}} probability.
Following \cite{Saon2019}, sequence noise mixed from up to 4 utterances is injected with 40\% probability and 0.3 weight.
The filterbank output is mean-and-variance normalized at the speaker level, and first ($\Delta$) and second ($\Delta\Delta$) derivatives are also calculated.
The final features presented to the network are also processed through a SpecAugment block that uses the \texttt{SM} policy~\cite{Park2019} with $p=0.3$ and no time warping.

\begin{figure}
\centering
\begin{subfigure}{0.31\linewidth}
\includegraphics[width=\linewidth]{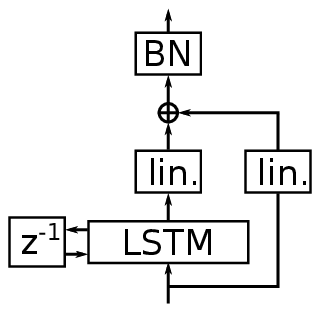}
\caption{} 
\label{fig:1a}
\end{subfigure}
\hspace{.1\linewidth}
\begin{subfigure}{0.5\linewidth}
\includegraphics[width=\linewidth]{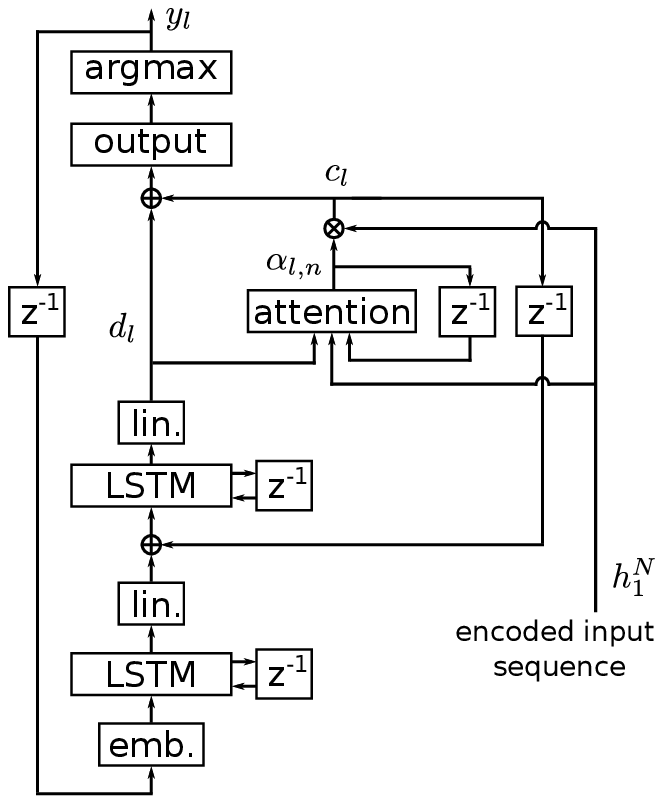}
\vspace{-7mm}
\caption{} 
\label{fig:1b}
\end{subfigure}
\vspace{-3mm}
\caption{{\label{fig:encdec}
{\it
(a) Building block of the encoder; (b) attention based decoder network used in the experiments.
}}}
\vspace{-6mm}
\end{figure}

The encoder network comprises 8 bidirectional LSTM layers with 1536 nodes per direction per layer \cite{Hochreiter97,Schuster}. As shown in Fig.~\ref{fig:1a}, each LSTM block in the encoder includes a residual connection with a linear transformation that bypasses the LSTM, a 1024-dimensional linear reduction layer on the LSTM output, and batch-normalization (BN) of the block output. A pyramidal structure~\cite{Chan2016} in the first two LSTM layers reduces the frame rate by a factor of 4.
The final dimension of the encoder output is 256, enforced by a linear bottleneck.
We apply 30\% dropout to the LSTM outputs and 30\% drop-connect to the hidden-to-hidden matrices~\cite{pmlr-v28-wan13,JMLR:v15:srivastava14a}. As suggested by \cite{Gal2016}, the weight dropout is fixed for a batch of sequences.

The attention based decoder model is illustrated in Fig.~\ref{fig:1b}.
The decoder models the sequence of 600 BPE units estimated on characters~\cite{subword-nmt}, where the BPE units are embedded in 256 dimensions.
We use additive, location aware attention, without key/value transformations, and the attention is smoothed by 256, 5-dimensional kernels~\cite{chorowski15}.
The decoder block consists of 2 unidirectional LSTM layers: one is a dedicated language-model-like component with 512 nodes that operates only on the embedded predicted symbol sequence, and the other is a 768 unit layer processing acoustic and symbol information.
The output of both LSTMs is reduced to 256 dimensions by a linear bottleneck~\cite{Vesely2011}.
Fixed sequence-level weight dropout of 15\% is applied in the decoder LSTMs, a dropout of 5\% is applied to the embeddings, and a dropout of 15\% is applied to the decoder LSTM outputs.
The second LSTM in the decoder also uses zoneout, where the cell state update is deactivated with 15\% probability and the recurrent feedback from the output maintains its previous value with 5\% probability.

Overall, the model has 280M parameters, of which only 5.4M are in the decoder.
Aiming at the best word error rate, this design choice is based on our observation that an external language model has significantly larger effect if the decoder is not over-parametrized~\cite{Tuske2019}.
The model is trained for 250 epochs on 32~P100 GPUs in less than 4 days using a PyTorch~\cite{pytorch} implementation of distributed synchronous SGD with up to 32 sequences per GPU per batch.
Training uses a learning rate of 0.03 and Nesterov momentum~\cite{nesterov1983} of 0.9.
The weight decay parameter is \mbox{4e-6}, the label smoothing parameter is 0.35, and teacher forcing is fixed to 0.8 throughout training.
In the first 3 epochs the learning rate is warmed up and batch size is gradually increased from 8 to 32~\cite{Goyal2017}.
In the first 35 epochs, the neural network is trained on sequences sorted in ascending order of length of the input.
Afterwards, batches are randomized within length buckets, ensuring that a batch always contains sequences with similar length.
Weight noise from a normal distribution with mean 0.0 and variance 0.015 is switched on after 70 epochs.
After 110 epochs, the updates of sufficient statistics in the batch-normalization layers are turned off, converting them into fixed affine transformations.
The learning rate is annealed by 0.9 per epoch after 180 epochs of training, and simultaneously label smoothing is also switched off.

The external language model (LM) is built on the BPE segmentation of 24M words from the Switchboard and Fisher corpora (from which the SWB-300 corpus roughly corresponds to 3M words).
It is trained for 40 epochs using label smoothing of 0.15 in the first 20 epochs.
The baseline LM has 57M parameters and consists of 2 unidirectional LSTM layers with 2048 nodes~\cite{Sundermeyer2012} trained with drop-connect and dropout probabilities of 15\%.
The embedding layer has 512 nodes, and the output of the last LSTM is projected to 128 dimensions.
When the LM is trained and evaluated across utterances, consecutive segments of a single-channel recording are grouped together up to 40 seconds.
Perplexities (PPL) are measured at the word level on the concatenation of ground truth transcripts, while the WER is obtained by retaining the LM state of the single-best hypothesis of the preceding utterance.

Decoding uses simple beam search with a beam width of 60 hypotheses and no lexical prefix tree constraint~\cite{Ney1992}.
The search performs shallow fusion of the encoder-decoder score, the external language model score, a length normalization term, and a coverage term~\cite{GulcehreFXCBLBS15,Hannun14,Tu2016}.
For more details, please refer to~\cite{Tuske2019}.
Hub5'00 is used as a development set to optimize decoding hyperparameters, while Hub5'01 and RT03  are used as final test sets.

\renewcommand{\baselinestretch}{0.93}\normalsize
\section{Experimental results}
Our current setup is the result of incremental development.
Keeping in mind that several other equally powerful setups probably exist, the focus of the following experiments is to investigate ours around the current optimum.

\subsection{Effect of data preparation}
We first investigate the importance of different data processing steps.
The \texttt{s5c} Kaldi recipe includes a duplicate filtering step, in which the maximum number of occurrences of utterances with the same content is limited. We measure the impact of duplicate filtering and also the effect of filtering out word fragments and noise tokens from the training transcripts.
Since the LM is trained on identically filtered transcripts from Fisher+Switchboard data, word fragment and noise token filters were applied consistently.
The results are summarized in Table~\ref{tab:preparation}.
Deactivating the duplicate filter is never harmful when an external LM is used, and the gains on CallHome can be substantial. Considering performance on the complete Hub5'00 data, the best systems either explicitly handle both word fragments and noise tokens or filter them all out. When an external LM is used, the best results are obtained when word fragment and noise token filters are activated and the duplicate filter is deactivated. This setting is also appealing in cases where the external LM may be trained on text data that will not contain word fragments or noise; thus, the remaining experiments are carried out with this system setting.

\begin{table}
  \centering
  \caption{Effect of data preparation steps on WER [\%] measured on Hub5'00, models are trained on SWB-300. The second row corresponds to the Kaldi \texttt{s5c} recipe.}
  \vspace{-3mm}
  \begin{tabular}{|c|c|c|c|c|c|c|}
    \hline
  \multicolumn{3}{|c}{filter} 
               & \multicolumn{2}{|c|}{w/o LM} & \multicolumn{2}{|c|}{w/ LM}              \\
\hline
frag. &  noise & dup.       &  swb & chm  & swb & chm  \\
\hline
\hline
         &          &           & {\bf 7.5}  & {\bf 14.3} & 6.7        & {\bf 12.6} \\  
\hline         
         &          & $\checkmark$  & 7.8         & 14.8        & 6.5        & 13.2   \\  
\hline         
         & $\checkmark$ &           & 7.6         & 15.1        & 6.5        & 13.1   \\  
\hline         
         & $\checkmark$ & $\checkmark$  & {\bf 7.5}  & 15.1        & 6.6        & 13.3   \\  
\hline         
$\checkmark$ & $\checkmark$ &           & 7.6         & 14.6        & {\bf 6.4} & 12.7  \\  
\hline         
$\checkmark$ & $\checkmark$ & $\checkmark$  & 7.7         & 14.7        & {\bf 6.4}        & 13.1   \\  
\hline
\end{tabular}                              
\label{tab:preparation}
\vspace{-3mm}
\end{table}

\subsection{Ablation study}
In a second set of experiments, we characterize the importance of each of the regularization methods described in Sec.~\ref{sec:methods} for our model performance by switching off one training method at a time without re-optimizing the remaining settings. In these experiments, decoding is performed without an external language model.
Curriculum learning is evaluated by either switching to randomized batches after 35 epochs or leaving the sorting on throughout training.
We also test the importance of $\Delta$ and $\Delta\Delta$ features~\cite{Furui1981}.
Sorting the results by decreasing number of absolute errors on Hub5'00, Table~\ref{tab:ablation} indicates that each regularization method contributes to the improved WER.
SpecAugment is by far the most important method, while using $\Delta$ and $\Delta\Delta$ features or switching off the curriculum learning in the later stage of training have marginal but positive effects.
Other direct input level perturbation steps (speed/tempo perturbation and sequence noise injection) are also key techniques that can be found in the upper half of the table.
If we compare the worst and baseline models, we find that the relative performance difference between them is nearly unchanged by including the external LM in decoding.
Without the LM, the gap is 18\% relative, while with the LM the gap is 17\% relative.
This clearly underlines the importance of the regularization techniques.

\begin{table}
  \centering
  \caption{Ablation study on the final training recipe, models are trained on SWB-300. WER is measured without using external LM.}
\vspace{-3mm}
  \begin{tabular}{|c|l|c|c|@{}c@{}|c|}
    \hline
  \multicolumn{2}{|c|}{}        &  \multicolumn{3}{c|}{WER [\%]} & \#err. \\  \cline{3-5}
  \multicolumn{2}{|c|}{}        &  swb  &  chm &\hspace{1mm}total\hspace{1mm} & [word] \\
\hline
\multirow{14}{*}{\rotatebox{90}{discarded ingredient}}  
 &  SpecAugment                  & 9.1   & 17.3 & 13.2 & 5665 \\
 &  speed/tempo pert.            & 8.2   & 15.5 & 11.9 & 5113 \\
 &  dropout                      & 8.0   & 15.6 & 11.8 & 5092 \\
 &  label smoothing              & 7.9   & 15.6 & 11.8 & 5072 \\
 &  sequence noise               & 7.8   & 15.6 & 11.7 & 5027 \\
 &  weight noise                 & 7.9   & 15.2 & 11.6 & 4973 \\
 &  weight decay                 & \bf{7.6}   & 15.1 & 11.4 & 4909 \\
 &  DropConnect                  & 7.7   & 15.1 & 11.4 & 4897 \\
 &  BN freezing                  & 7.8   & 14.9 & 11.4 & 4880 \\
 &  scheduled samp.              & 7.7   & 14.8 & 11.3 & 4856 \\
 &  zoneout (in dec.)            & 7.7   & 14.9 & 11.2 & 4831 \\
 &  residual (in enc.)           & 7.7   & 14.7 & 11.2 & 4827 \\
 &  random. batch                & \bf{7.6}   & 14.7 & 11.2 & 4794 \\
 &  +$\Delta$, +$\Delta\Delta$   & 7.7   & \bf{14.6} & \bf 11.1 & 4792 \\
\hline 
  \multicolumn{1}{|c}{} & baseline & \bf{7.6}   & \bf{14.6} & \bf 11.1 & \bf{4775} \\
\hline
\end{tabular}
\label{tab:ablation}
\vspace{-2mm}
\end{table}

\subsection{Optimizing the language model}
\label{sec:optlm}
The following experiments summarize our optimization of the LM.
Compared to our previous LM~\cite{Saon2019}, we measure better perplexity and WER if no bottleneck is used before the softmax layer (rows 1 and 3 in Table~\ref{tab:lmopt}).
Increasing the model capacity to 122M parameters results in a significant gain in PPL only after the dropout rates are tuned (rows 3, 5 and 6).
Similar to \cite{Tuske2018,Xiong2018}, significant PPL gain is observed if the LM was trained across utterances.
However, this PPL improvement does not translate into reduced WER with a bigger model when cross utterance modeling is used (rows 4 and 7).
Thus, in all other experiments we use the smaller, 57M-parameter model.

\begin{table}
  \centering
  \caption{Optimizing dropout (dropo.), DropConnect (dropc.), layer and bottleneck (bn) size for LSTM LM, optionally modeling across utterances (x-utt.). WER is measured in shallow fusion with the best SWB-300 seq2seq ASR model.}
\vspace{-3mm}
  \begin{tabular}{|@{}c@{}|@{}c@{}|@{}c@{}|@{}c@{}|@{}c@{}|@{}c@{}|@{}c@{}|@{}c@{}|@{}c@{}|}
\hline
 \multicolumn{5}{|c|}{model}               & \multicolumn{2}{c|}{PPL$_{\text{word}}$} & \multicolumn{2}{c|}{WER} \\
\hline
 \hspace{1mm}dropo.\hspace{1mm} & \hspace{1mm}dropc.\hspace{1mm} & \hspace{1mm}width\hspace{1mm}  & \hspace{1.5mm}bn\hspace{1.5mm}  &  x-utt. & \hspace{1.5mm}CV\hspace{1.5mm}   & \hspace{1mm}Hub5'00\hspace{1mm} & \hspace{1mm}swb\hspace{1mm} & \hspace{1mm}chm\hspace{1mm}  \\
\hline
\multirow{5}{*}{15\%} & \multirow{5}{*}{15\%} & \multirow{4}{*}{2048}  & \multirow{2}{*}{128} &          & 56.7 & 65.7 & 6.7 & 13.2 \\ \cline{5-9}
          &            &        &     & $\checkmark$ & 46.3 & 52.7 & 6.5 & 13.1 \\ \cline{4-9}
          &            &        & \multirow{5}{*}{-}  &          & 52.9 & 61.2 & 6.6 & 13.2  \\  \cline{5-9}
          &            &        &     & $\checkmark$ & 44.1 &  50.1 & 6.4 & {\bf 12.7} \\ \cline{3-3}\cline{5-9}
          &            & \multirow{3}{*}{3072} &     &          & 53.9 & 64.0 &  6.7   & 13.2     \\ \cline{1-2}\cline{6-9}
 \multirow{2}{*}{30\%} & \multirow{2}{*}{30\%} &  &   &          & 50.3 & 58.3 &  6.4   & 13.1     \\ \cline{5-9}
                       &                       &  &   & $\checkmark$ & {\bf 41.4} & {\bf 47.0}  & {\bf 6.3} & 12.8 \\
\hline
\end{tabular}
\label{tab:lmopt}
\vspace{-3mm}
\end{table}

\subsection{Effect of beam size and number of parameters}
A 280M-parameter model may be larger than is practical in many applications. Thus, we also conduct experiments to see if this model size is necessary for reasonable ASR performance.
Models are trained without changing the training configuration, except that the size or number of LSTM layers is reduced.
As Table~\ref{tab:modelsize} shows, although our smallest attention based model achieves reasonable results on this task, a significant loss is indeed observed with decreasing model size, especially on CallHome.
Nevertheless, an external language model reduces the performance gap.
A small, 57M-parameter model together with a similar size language model is only 5\% relative worse than our largest model.
We note that this model already outperforms the best published attention based seq2seq model \cite{Park2019}, with roughly 66\% fewer parameters.

\begin{table}
  \centering
  \caption{Effect of model size, models are trained on SWB-300.}
\vspace{-3mm}
  \begin{tabular}{|@{}c@{}|@{}c@{}|@{}c@{}|@{}c@{}|@{}c@{}||@{}c@{}|@{}c@{}|@{}c@{}|@{}c@{}|}
    \hline
\multicolumn{3}{|c|}{enc.}   & dec.   &          & \multicolumn{4}{c|}{WER}   \\ \cline{1-4} \cline{6-9}
\multirow{2}{*}{\eightpt \hspace{0.6mm}depth\hspace{0.6mm}}        & \multirow{2}{*}{\eightpt \hspace{0.6mm}LSTM\hspace{0.6mm}}  & \multirow{2}{*}{\eightpt lin.}     & \multirow{2}{*}{\eightpt \hspace{0.6mm}LSTM\hspace{0.6mm}}   & \#par. & \multicolumn{2}{c|}{\eightpt w/o LM} & \multicolumn{2}{c|}{\eightpt w/ LM}   \\  \cline{6-9}
        &        &      &        &  [M]  &  swb  &  chm &  swb  &  chm \\
\hline        
\multirow{2}{*}{6}
        & \textcolor{white}{0}512   & \textcolor{white}{0}384  & 512    & \textcolor{white}{0}28.5 & \hspace{1mm} 9.4 \hspace{1mm} & \hspace{1mm} 17.0 \hspace{1mm} & \hspace{1mm} 7.4 \hspace{1mm} & \hspace{1mm} 14.6 \hspace{1mm} \\ \cline{2-9}
        & \multirow{2}{*}{\textcolor{white}{0}768}   & \multirow{2}{*}{\textcolor{white}{0}512}  & \multirow{5}{*}{768}
                                 & \textcolor{white}{0}57.3 &  8.3         & 15.5        &  6.6        & 13.4 \\ \cline{1-1} \cline{5-9}
\multirow{4}{*}{8}
        &                           &                          &        & \textcolor{white}{0}75.1 &  8.4  & 15.1 &  6.8  & 13.4 \\ \cline{2-3} \cline{5-9}
        & 1024   & \textcolor{white}{0}640  &        & 125.0 &  7.6  & 15.2 &  6.5  & 13.3 \\ \cline{2-3} \cline{5-9}
        & 1280   & \textcolor{white}{0}896  &        & 201.6 &  {\bf 7.5}  & 15.0        &  {\bf 6.4} & 12.9 \\ \cline{2-3} \cline{5-9}
        & 1536   & \hspace{0.2mm}1024\hspace{0.2mm} &        & \hspace{0.6mm}280.1\hspace{0.6mm} &  7.6         & {\bf 14.6} &  {\bf 6.4} & {\bf 12.7} \\
\hline
\end{tabular}
\label{tab:modelsize}
\end{table}

\begin{figure}
\vspace{-1mm}
\centering
\hspace{-1mm}\includegraphics[width=0.97\linewidth]{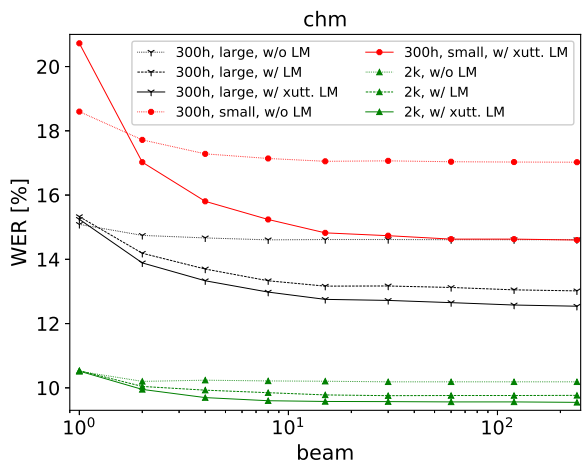}
\vspace{-4mm}
\caption{{\label{fig:beamchm}
{\it
Effect of beam size on word error rate (WER) measured on the CallHome (chm) part of Hub5'2000. ``300h'' indicates models trained on SWB-300, whereas ``2k'' corresponds to the 2000-hour Switchboard+Fisher training setup.
}}}
\vspace{-2mm}
\end{figure}

Additional experiments are carried out to characterize the search and modeling errors in decoding. The results of tuning the beam size and keeping the other search hyperparameters unchanged are shown in Fig.~\ref{fig:beamchm}.
``Small'' denotes the 57M model, while ``large'' denotes the 280M model.
When greedy search (beam 1) is used, the external language model increases WER, an effect that might be mitigated with re-optimized hyperparameters.
Nevertheless, if a beam of at least 2 hypotheses is used, the positive effect of the language model is clear.
We also observe that without the language model the search saturates much earlier, around beam 8, fluctuating within only a few absolute errors afterwards.
On the contrary, decoding with the language model, we measure consistent but small gains with larger beams.
The minimum number of word errors was measured with a relatively large beam of 240.
The figure also shows that the effect of a cross-utterance language model grows with larger beams.
Lastly, if the model is trained on 2000 hours of speech data (see next section), the extremely fast greedy decoding gives remarkably good performance.
Although the importance of beam search decreases with an increased amount of training data, we still measure 10\% relative degradation compared to a system with a cross-utterance LM and wide (240) beam search.

\subsection{Experiments on Switchboard-2000}
\vspace{-.15mm}
As a contrast to our best results on Switchboard-300, we also train two seq2seq models on the 2000-hour Switchboard+Fisher data.
The first model consists of 10 encoder layers, has 360M parameters, and is trained for only 50 epochs.
The second model (\texttt{lrg}) has 660M parameters, 14 encoder and 4 decoder layers, and training runs for 250 epochs.
Our overall results on the Hub5'00 and other evaluation sets are summarized in Table~\ref{tab:overall}.
The results in Fig.~\ref{fig:beamchm} and Table~\ref{tab:overall} show that adding more training data and using a fairly large model greatly improves the system, over 30\%~relative in some cases.
For comparison with others, our best 2000-hour system reaches 7.1\% and 6.0\% WER on rt02 and rt04.
It is also worth to note that human performance is at 6.0\%, 4.5\%, 4.7\% WER on the rt0\{2,3,4\} sets, measured as in \cite{Saon2017}.
We observe that the regularization techniques, which are extremely important on the 300h setup, are also beneficial to train much larger models using big data.
Considering the recognition speed and applicability of the \texttt{2k-lrg} model without using external language model, we measure 0.73~--~0.77 real-time factor and 6.5~--~6.4\% total WER on Hub5'00 after varying the beam between 4 and 16, using a single core of an Intel Xeon Platinum 8280 processor and 8-bit integer weight quantization.

\begin{table}
  \centering
  \caption{Detailed results with the best performing systems on both SWB-300 and SWB-2000.}
\vspace{-3mm}
  \begin{tabular}{|@{}c@{}||@{}c@{}|@{}c@{}||@{}c@{}|@{}c@{}||@{}c@{}|@{}c@{}|@{}c@{}||@{}c@{}|@{}c@{}|}
    \hline
\multicolumn{1}{|@{}c@{}||}{} & \multicolumn{1}{@{}c@{}|}{\multirow{3}{*}{\hspace{0.5mm}\parbox[c]{5mm}{ext.\\LM}\hspace{0.5mm}}}  &  & 
                                            \multicolumn{2}{@{}c@{}||}{hub5'00}  & \multicolumn{3}{@{}c@{}||}{hub5'01}  & \multicolumn{2}{@{}c@{}|}{rt03} \\ \cline{4-10}
\multicolumn{1}{|@{}c@{}||}{SWB} & & \hspace{1.0mm}xutt.\hspace{1.0mm} & \multirow{2}{*}{\hspace{1.0mm}swb\hspace{1.0mm}} & \multirow{2}{*}{\hspace{1.0mm}chm\hspace{1.0mm}} & \multirow{2}{*}{\hspace{1.0mm}swb\hspace{1.0mm}} & \hspace{1.0mm}swb2\hspace{1.0mm} & \hspace{1.0mm}swb2\hspace{1.0mm} & \multirow{2}{*}{\hspace{1.0mm}swb\hspace{1.0mm}} & \multirow{2}{*}{\hspace{1.0mm}fsh\hspace{1.0mm}} \\
\multicolumn{1}{|@{}c@{}||}{} &                             &          &      &       &      &  p3  &  p4  &      &     \\
\hline
\hline
\multirow{3}{*}{\rotatebox{0}{300}}   
         &                      &          & 7.6  &  14.6 &  8.1 & 11.0                    &  15.7  & 17.8 & 10.5                    \\ \cline{2-10}  
         &       $\checkmark$       &          & 6.5  &  13.0 &  7.0 & \textcolor{white}{0}9.3 &  13.8  & 15.0 & \textcolor{white}{0}8.8 \\ \cline{2-10} 
         &       $\checkmark$       & $\checkmark$ & \bf{6.4}  &  \bf{12.5} &  \bf{6.8} & \textcolor{white}{0}\bf{9.1} &  \bf{13.4}  & \bf{14.8} & \textcolor{white}{0}\bf{8.4} \\  
\hline       
\hline       
\multirow{3}{*}{\hspace{1mm}2k\hspace{1mm}}
         &                      &          & 5.9  &  10.2 &  \textcolor{white}{}6.8  &   \textcolor{white}{0}8.7  &  11.7  & 10.4                      & \textcolor{white}{0}7.2    \\ \cline{2-10}
         &       $\checkmark$       &          & \bf{5.5}  & \textcolor{white}{0}9.8 &  \textcolor{white}{}6.6  &   \textcolor{white}{0}8.3  &  11.5  & \textcolor{white}{0}\bf{9.8}   & \textcolor{white}{0}6.7    \\ \cline{2-10}
         &       $\checkmark$       & $\checkmark$ & 5.6  & \textcolor{white}{0}\bf{9.5} &  \textcolor{white}{}\bf{6.5}  & \textcolor{white}{0}\bf{8.2}  &  \bf{11.4}  & \textcolor{white}{0}\bf{9.8}   & \textcolor{white}{0}\bf{6.6}    \\ 
\hline
\hline
\multirow{2}{*}{\hspace{.4mm}2k-lrg\hspace{.4mm}}
         &                          &              &     4.8  & \textcolor{white}{0}{8.0} &  \textcolor{white}{}{5.5}  & \textcolor{white}{0}{6.7}  &  {10.3}  & \textcolor{white}{0}{8.5}   & \textcolor{white}{0}{7.0}    \\ \cline{2-10}           
         &       $\checkmark$       & $\checkmark$ & \bf{4.7} & \textcolor{white}{0}\bf{7.8} &  \textcolor{white}{}\bf{5.2}  & \textcolor{white}{0}\bf{6.3}  & \textcolor{white}{0}\bf{9.7}  & \textcolor{white}{0}\bf{8.2}   & \textcolor{white}{0}\bf{6.4}    \\ 
\hline
\end{tabular}
\label{tab:overall}
\vspace{-3mm}
\end{table}

\vspace{-.5mm}
\section{Comparison with the literature}
\vspace{-1mm}
For comparison with results in the literature we refer to the Switchboard-300 results in \cite{Park2019,Irie2019asru,Hadian2018,Audhkhasi2019} and the \mbox{Switchboard-2000} results in \cite{Xiong2018,Hadian2018,Saon2017,Battenberg2017,Kurata2017,Nguyen2019,capio}.
Our 300-hour model not only outperforms the previous best attention based encoder-decoder model~\cite{Park2019} by a large margin, it also surpasses the best hybrid systems with multiple LMs~\cite{Irie2019asru}.
Our single system result on Switchboard-2000 is also better than the best system combination results reported to date.

\section{Conclusions}
\vspace{-1.0mm}
We presented an attention based encoder-decoder setup which achieves state-of-the-art performance on both Switchboard 300 and 2000.
A rather simple model built from LSTM layers and a decoder with a single-headed attention mechanism outperforms the standard hybrid approach.
This is particularly remarkable given that in our model neither a pronunciation lexicon nor a speech model with explicit hidden state representations is needed.
We also demonstrated that excellent results are possible with smaller models and with practically search-free, greedy decoding.
The best results were achieved with a speaker independent model in a single decoding pass, using a minimalistic search algorithm, and without any attention mechanism in the language model.
Thus, we believe that further improvements are still possible if we apply a more complicated sequence-level training criterion and speaker adaptation.
As a further possible extension of this study, the training of conventional ASR models should also be revisited for fairer comparison of different modeling approaches.

\ninept
\bibliographystyle{IEEEtran}
\bibliography{mybib}

\end{document}